\documentclass{Interspeech}

\interspeechcameraready

\title{AISHELL-5: The First Open-Source In-Car Multi-Channel Multi-Speaker Speech Dataset for Automatic Speech Diarization and Recognition}
\vspace{-3em}
\author[affiliation={1}]{Yuhang}{Dai}
\author[affiliation={1}]{He}{Wang}
\author[affiliation={1}]{Xingchen}{Li}
\author[affiliation={1}]{Zihan}{Zhang}
\author[affiliation={1}]{Shuiyuan}{Wang}
\author[affiliation={1*}]{Lei}{Xie}
\author[affiliation={2}]{Xin}{Xu}
\author[affiliation={2}]{Hongxiao}{Guo}
\author[affiliation={2}]{Shaoji}{Zhang}
\author[affiliation={2}]{Hui}{Bu}
\author[affiliation={3}]{Wei}{Chen}
\affiliation{}{Audio, Speech and Language Processing Group(ASLP@NPU)}{School of Computer Science, \\ Northwestern Polytechnical University, Xi'an, China}
\affiliation{}{Beijing AISHELL Technology Co., Ltd.}{Beijing, China}
\affiliation{}{Li Auto Inc.}{Beijing, China}
\email{yhdai@mail.nwpu.edu.cn, lxie@nwpu.edu.cn}


\usepackage{comment}
\usepackage{multirow}
\usepackage{float}
\usepackage{graphicx}
\usepackage{svg}
\usepackage{subcaption}
\usepackage{pifont}
\usepackage{authblk}

\begin{document}

\maketitle

\renewcommand{\thefootnote}{\fnsymbol{footnote}}
\footnotetext[1]{Corresponding author.}
\renewcommand{\thefootnote}{\arabic{footnote}}

\begin{abstract}
This paper delineates AISHELL-5, the first open-source in-car multi-channel multi-speaker Mandarin automatic speech recognition (ASR) dataset. 
AISHLL-5 includes two parts: 
(1) over 100 hours of multi-channel speech data recorded in an electric vehicle across more than 60 real driving scenarios.
This audio data consists of four far-field speech signals captured by microphones located on each car door, as well as near-field signals obtained from high-fidelity headset microphones worn by each speaker.
(2) a collection of 40 hours of real-world environmental noise recordings, which supports the in-car speech data simulation.
Moreover, we also provide an open-access, reproducible baseline system based on this dataset.
This system features a speech frontend model that employs speech source separation to extract each speaker's clean speech from the far-field signals, along with a speech recognition module that accurately transcribes the content of each individual speaker.
Experimental results demonstrate the challenges faced by various mainstream ASR models when evaluated on the AISHELL-5.
We firmly believe the AISHELL-5 dataset will significantly advance the research on ASR systems under complex driving scenarios by establishing the first publicly available in-car ASR benchmark.

\end{abstract}

\renewcommand{\arraystretch}{1.3}
\begin{table*}[ht]
\centering

\caption{An illustration of the diverse sub-scenes in AISHELL-5, including scene numbers (day and night), window status, car driving status, air-conditioning status, and car stereo status. Apart from the sub-scene shown in the table, number 1 represents the day scene, while number 2 denotes the night scene. For example, A1 indicates a stopped car with both the air-conditioning and car stereo turned off during the day scene.}
    \begin{tabular}{cccccccc}
    \hline
    \multicolumn{4}{c}{N: Window open}                           & \multicolumn{4}{c}{M: All windows are closed} \\ \hline
    \multicolumn{4}{c}{Window state}                             & \multicolumn{4}{c}{Car state}                 \\ \hline
    Index & Driver's side window & \multicolumn{2}{c}{Sunroof}  & Index  & Drive state  & AC      & Car Stereo \\ \hline
    N1     & Open 1/3             & \multicolumn{2}{c}{Closed}   & A       & Stopped      & Off     & Off        \\ 
    N2     & Closed               & \multicolumn{2}{c}{Open 1/2} & B       & Stopped      & Medium  & Off        \\
    N3     & Open 1/2             & \multicolumn{2}{c}{Open 1/2} & C       & Stopped      & High    & Medium     \\ \cline{1-4}
    \multicolumn{4}{c}{\multirow{2}{*}{Car state}}               & D       & 0-40 km/h    & Off     & Off        \\
    \multicolumn{4}{c}{}                                         & E       & 0-40 km/h    & Medium  & Off        \\ \cline{1-4}
    Index & Drive state          & AC          & Car Stereo     & F       & 0-40 km/h    & High    & Medium     \\ \cline{1-4}
    A      & Stopped              & Off         & Off            & G       & 40-80 km/h   & Off     & Off        \\
    B      & Stopped              & Medium      & Off            & H       & 40-80 km/h   & Medium  & Off        \\
    C      & Stopped              & High        & Medium         & I       & 40-80 km/h   & High    & Medium     \\
    D      & 0-60 km/h            & Off         & Off            & J       & 80-120 km/h  & Off     & Off        \\
    E      & 0-60 km/h            & Medium      & Off            & K       & 80-120 km/h  & Medium  & Off        \\
    F      & 0-60 km/h            & High        & Medium         & L       & 80-120 km/h  & High    & Medium     \\ \hline
           
    \end{tabular}
\label{scene}
\end{table*}

\begin{figure*}[ht]
    \centering
    \includegraphics[clip, trim=7cm 12cm 3.5cm 2cm, width=1.4\linewidth]{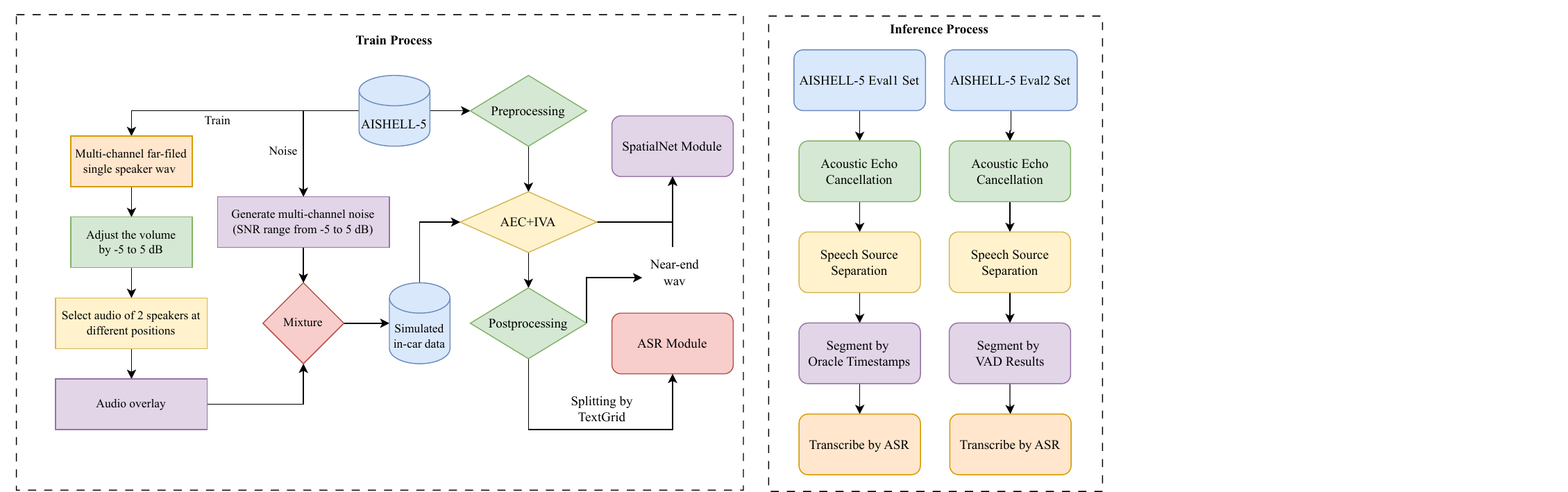}  % 替换为你的图像文件路径
    \caption{Structure of our baseline system, including train process and inference process}
    \label{fig:enter-label}
\end{figure*}

\section{Introduction}
% Unlike common scenarios for automatic speech recognition (ASR) systems, such as home or meeting, the acoustic environment of the closed and irregular cockpit is complex.
% Unlike typical automatic speech recognition (ASR) applications in homes or meetings, in-car environments present unique challenges. 
Unlike common automatic speech recognition (ASR) applications in the home or smart assistant scenarios, in-car ASR systems encounter a range of unique challenges.
% The accuracy of in-car ASR is significantly degraded by various internal and external noises, including wind, engine, tire, car audio, and road sounds.
These systems must contend with various internal and external noise sources, including wind, engine sound, tire noise, car stereos, and nearby vehicles, all of which contribute to a highly complex acoustic environment.
% In most cases, in-car ASR systems fight with various internal and external sources of noise, including wind, engine, tires, car stereos, and other vehicles around, combined with irregular cockpits, leading to an extremely complex acoustic environment.
Moreover, conversations between the driver and passengers frequently introduce overlapping speech, further complicating the recognition progress.
% This complexity necessitates advanced in-car speech processing techniques that integrate speech frontend processing with speech source separation and ASR modules. 
% To achieve good results in speech recognition of in-car scenarios, there are two key problems:  First, the complex and diverse acoustic environment of the in-car scenarios; second, the frequent overlapping speech during conversations among multiple speakers.
% Therefore, we need not only a highly robust acoustic model for in-car scenarios but also data that is specifically designed for these scenarios to deal with overlapping speech problem.
% Basically, building a robust in-car speech recognition system hinges on addressing two key issues: the complex acoustic environment in the irregular cockpit, and the prevalence of overlapping speech. 
Fundamentally, building a robust in-car speech recognition system hinges on addressing two key issues: the complex acoustic environment, and the frequent occurrence of overlapping speech.
While innovative designs for speech frontend and recognition model architectures are crucial, there is an increasing demand among researchers for specialized open-source datasets that can facilitate in-car data augmentation and model training.
% In addition to designing ingenious speech frontend or recognition model structures, researchers are increasingly calling for more specialized open-source datasets that can be used for in-car data augmentation and model training.
% Therefore, progress requires not only robust acoustic model design tailored for various in-car noise signals and speaker overlap interruption but also specialized datasets designed to cover in-car driving scenarios and speaker overlap situations.

% First, regarding the issue of speaker overlap, it has attracted the attention of many researchers, and many existing open-source datasets are also designed to alleviate this problem.
% AISHELL-4~\cite{fu2021aishell} is an eight-channel Mandarin meeting scene speech dataset recorded by microphone arrays, encompassing key characteristics of real meeting scenarios such as pauses, overlaps, speaker turns, and noise. 
% It provides a good solution for the overlapping speech problem and advances research in this area. However, it is not designed for in-car scenarios. 
First, the issue of speaker overlap has attracted a lot of attention from researchers, leading to the development of several open-source datasets specifically designed to cover this situation.
AISHELL-4~\cite{fu2021aishell} and AliMeeting~\cite{Yu2022M2Met} are both multi-speaker Mandarin ASR datasets recorded by multi-channel microphone arrays in indoor meeting scenarios, of which the speech captures lots of key characteristics of free-talk conversation, such as pauses and speaker overlaps. 
% So far, these two datasets remain important benchmarks for testing ASR performance in multi-speaker Mandarin meeting scenarios, and they severely test the model's robustness to far-field and overlapped speech signals. 
To date, these two datasets have played a crucial role in evaluating ASR performance in multi-speaker Mandarin meeting scenarios, effectively testing the robustness of models against far-field and overlapping speech.
% LibriMix~\cite{cosentino2020librimix}, derived from LibriSpeech~\cite{panayotov2015librispeech}, with a large data scale, offers higher generalization ability for model training and is used for noisy speech separation tasks. 
LibriMix~\cite{cosentino2020librimix} is an open-source English dataset designed for speech separation and multi-speaker ASR in noisy environments. 
It derives from the freely available LibriSpeech~\cite{panayotov2015librispeech} dataset and WHAM!~\cite{wichern2019wham} noise dataset, where LibriSpeech provides clean speech signals and WHAM! provides noise samples to create noisy mixtures. 
However, the data in the LibriMix dataset, which is constructed by directly splicing and overlapping existing speech samples, cannot accurately reflect real-world scenarios.
Except for the open-source datasets, some researches~\cite{Simulation-ravanelli2017realistic, Simulation-wang2024effective} bring up various data simulation methods, which alleviate the issue of low model accuracy caused by data shortage to a certain extent.
% These approaches fail to provide a satisfactory solution for the multi-channel, multi-scenario, and multi-speaker overlapping speech issues about in-car data.

% Second, in terms of the complex in-car acoustic environment, MagicData's released MDT-ASR-C001\footnote{https://www.magicdatatech.cn/datasets/asr/mdt-asr-c001-mandarin-chinese-speech-recognition-corpus} focuses on specific scenarios but is too small in scale and only contains single-speaker speech data, failing to cover the complexity of real-world in-car scenarios.
Second, regarding the complex in-car acoustic environment, only a few open-source datasets can cover it. 
MDT-ASR-C001\footnote{https://www.magicdatatech.cn/datasets/asr/mdt-asr-c001-mandarin-chinese-speech-recognition-corpus} is recorded in real in-car environments, reflecting real acoustic conditions and background noise. 
However, it comprises only 6 hours of data, which is insufficient to cover the complexities of real-world in-car driving scenarios.
%The ISCSLP 2022 Intelligent Cockpit Speech Recognition Challenge (ICSRC) held in 2022 released approximately 20 hours of validation and test sets, with data recorded using high-fidelity microphones in new energy vehicles.
The Intelligent Cockpit Speech Recognition Challenge (ICSRC) 2022~\cite{zhang2022iscslp} was successfully held and introduced a 20-hour single-channel in-car ASR test set, recorded by a high-fidelity microphone in a hybrid electric vehicle, focusing on the speech command recognition within smart cockpits.
% and provides valuable resources for the public test platform, which advances research about in-car speech recognition and intelligent cockpit voice command recognition. 
% However, due to the short duration, limited scenarios, and lack of multi-speaker conversational, the dataset has some limitations. 
Although the ICSRC dataset is indeed pioneering in the field of in-car ASR, its limited data size and focus on vehicle control commands restrict its applicability for enhancing ASR systems designed for multi-speaker dialogues in diverse driving conditions.
% The In-Car Multi-Channel Automatic Speech Recognition challenge(ICMC-ASR)~\cite{icmc} held in 2024 focuses on the field of speech processing and recognition under complex driving conditions.
In 2024, the In-Car Multi-Channel Automatic Speech Recognition (ICMC-ASR) challenge~\cite{icmc} was launched, attracting nearly 100 participating teams and focusing on advanced in-car speech processing under complex driving scenarios.
% The ICMC-ASR challenge provides over a hundred hours of data, including training sets, validation sets, noise set and test sets for two tracks. 
% It set two tracks: Track I is purely for ASR, named ASR track, where participants receive the ground-truth timestamps, indicating each speaker speaks at what time, in the evaluation set~\textbf{(Eval1)}, with character error rate (CER) as the evaluation metric;
The challenge consists of two tracks: Track I, designated as the ASR track, provides the ground-truth timestamps indicating when each speaker speaks. Its evaluation set~\textbf{(Eval1)} is measured by character error rate (CER) as the evaluation metric;
Track II is for automatic speech diarization and recognition (ASDR), with no timestamps provided in the evaluation set~\textbf{(Eval2)}, which requires a system to do speaker diarization first to get predicted timestamps of every speaker, and then transcribe their speech separately, with concatenated minimum permutation character error rate (cpCER) as the evaluation metric. 
%Kimi: Therefore, the ICMC-ASR data holds certain value for addressing the two issues in in-car speech processing technology research. 
% Therefore, the ICMC-ASR data is valuable for the two issues above about in-car speech processing techniques' research. 

% As mentioned above, in-car ASR systems eagerly require not only robust acoustic model design tailored for various in-car noise signals and speaker overlap interruption but also specialized datasets designed to cover in-car driving scenarios and speaker overlap situations.
As mentioned above, to enhance the accuracy and robustness of in-car ASR systems, there is an urgent need for a sizable open-source dataset that can simultaneously encompass speaker overlapping situations and different acoustic environments under various driving scenarios.
% To further promote research about in-car speech techniques, we have fixed the data-related issues of the dataset used in the ICMC challenge and now officially open-source it called AISHELL-5\footnote{http://www.aishelltech.com/aishell\_5}.
To further promote research on in-car speech processing, we fixed all the data-related issues of the ICMC-ASR challenge dataset, including audio truncation and mismatched transcription, and now officially open-source it, called AISHELL-5\footnote{https://www.aishelltech.com/aishell\_5}.
%Kimi: AISHELL-5 features multi-channel, multi-speaker, free-dialogue data with a large scale and about 60 different scenarios, covering most in-car acoustic environments. AISHELL-5 holds significant value for the field of in-car multi-channel speech processing. 
It features multi-channel, multi-speaker free-talking, as close as possible to real-world scenarios, with an over 100-hour scale.
In particular, it is recorded under 60 driving scenarios by varying lots of factors that may alter the in-car acoustic environment, including the driving speed, car window, car stereo, air-conditioning, driving day or night, and so on. 
% At the same time, we also open-source 40 hours of real-recorded noise data for data simulation research.
Moreover, 40 hours of real-recorded multi-channel noise signals are also open-sourced for promoting in-car speech data simulation research.
% AISHELL-5 primarily focus on the two issues: acoustic environment and overlaps about in-car speech processing scenarios, using \textbf{Eval1} and \textbf{Eval2} as the evaluation sets. 
% We also provide a baseline system for AISHELL-5 to promote research in this complex field. 
Based on the AISHELL-5 dataset, we also provide a baseline system that includes a frontend incorporating speech source separation and an ASR module.
Furthermore, we adopt the track settings and the evaluation sets from the ICMC-ASR challenge.
We believe that AISHELL-5, as a real-recorded and high-quality in-car speech dataset, can offer a certain amount of support for this increasingly important ASR application scenario, and contribute to advancing human-vehicle interaction towards greater accuracy and convenience.

\section{Dataset}
% 第一段说录制设置（录制用车、远场麦克风位置数量（4）、近场麦克风位置数量（each speaker）、说话人数量 with balanced gender、录制形式（每个session的说话人数量2 - 4人，free-talk的形式without restrict topic, ensure 交流的自然性和真实性））
% 第二段重点说场景 As we known 开头先说驾驶时会有哪些影响ASR准确率的因素（自然界噪声（风噪->窗户）、车本身的噪声：音乐播放器，空调，胎噪、other vihcles->车周围的环境噪声（白天黑夜、高速、街道）取决于周围车的数量和速度）？ 为了尽可能覆盖这些情况/场景，我们 carefully design了。
% Table 2修改 GT SD 和 Transcription  换成 音频条数 + 说话人数量
% 第三段 围绕Table 2去说。 先总说 Overall 我们收集了多少有效数据。再详细说 复述Table2。分成哪些集合+Noise data+时长+Session+Speaker Number。

% 说话人数量待确定
The AISHELL-5 dataset is recorded inside a hybrid electric car, with a far-field microphone placed above the door handles of all four doors to capture far-field audio from different areas of the car. 
Additionally, each speaker wears a high-fidelity microphone to collect near-field audio for data annotation. 
A total of 260 participants are involved in the recording with no notable accents. During the recording, 2-4 speakers are randomly seated in the four positions inside the car and engaged in free conversations without content restrictions to ensure the naturalness and authenticity of the audio data. The average duration of each session is 10 minutes.
The scripts for all our speech data are prepared in TextGrid format. 
Each session's TextGrid contains information such as the session duration, speaker details (number of speakers, speaker IDs, gender, etc.), timestamps for each audio segment, and the transcribed text.

In normal driving scenarios, the car typically contains various noises from both inside and outside. 
External noises include environmental sounds, wind noise, tire noise, etc., while internal noises come from sources like music players and air conditioning. 
These noises significantly impact the accuracy of in-car speech recognition systems. 
To comprehensively cover the various noise types encountered in real-world in-car scenarios, we carefully design the recording scenes. 
For environmental noise, recordings are made with different driving segments (urban streets and highways) during both daytime and nighttime. 
About the wind-induced noise and tire noise, we control the degree to which the car windows are open (fully closed, half open, and one-third open) and the car's speed (stationary, low-speed, medium-speed, and high-speed). 
For noise inside the car, we set the music player and air conditioning to different levels to cover a variety of in-car conditions. 
These different sub-scenes are numbered, and all sub-scenes are combined in various ways to form the final recording scenarios, resulting in over 60 recording scenarios in total. Specific settings for the sub-scenes are shown in Table\ref{scene}.

Overall, the AISHELL-5 dataset contains more than 100 hours of speech data, divided into 94 hours of training data (Train), 3.3 hours of validation data (Dev), and two test sets (Eval1 and Eval2), with durations of 3.3 and 3.58 hours. 
Each dataset includes far-field audio from 4 channels, with only the training set containing near-field audio. 
Additionally, to promote research on speech simulation techniques, we also provide a large-scale noise dataset (Noise), which has the same recording settings as the far-field data but without any speaker speech, lasting approximately 40 hours. Detailed information about the subsets of AISHELL-5 is provided in Table\ref{AISHELL_5_datasets}.

\begin{table}[ht]
\centering
\caption{Statistics about AISHELL-5 datasets, including the duration of segmented near-field audio (Duration), number of sessions (Session), number of speakers (Speaker) and including near-field audio or not (Near-field).}
\small
\setlength{\tabcolsep}{2pt} % 调整列间距
\renewcommand{\arraystretch}{1.2} % 增加行间距为1.2倍
\begin{tabular}{lccccc}
\hline
\textbf{Dataset} & \textbf{Duration (h)} & \textbf{Session} & \textbf{Speaker} & \textbf{Near-field} \\ \hline
Train & 94.75 & 568 & 147 & \ding{51} \\
Dev & 3.33 & 18 & 6 & \ding{55} \\
Eval1 & 3.30 & 18 & 6 & \ding{55} \\
Eval2 & 3.58 & 18 & 6 & \ding{55} \\
Noise & 40.29 & 60 & - & - \\ \hline
\end{tabular}
\label{AISHELL_5_datasets}
\end{table}

\begin{table*}[ht]
    \centering
    \caption{The results of our system on Eval1(CER(\%)) and Eval2(cpCER(\%)), using AEC + IVA and Spatialnet frontend with different ASR Models. The training data of AISHELL-5 includes far-field and near-field data with a total amount of 190 hours. (* indicates the model is fine-tuned with the AISHELL-5 training Set.)}
    \label{eval_results}
    \setlength{\tabcolsep}{4.5pt} % 调整列间距
    \renewcommand{\arraystretch}{1.2} % 增加行间距为1.2倍
\begin{tabular}{cccccccc}
\hline
\multirow{2}{*}{Model Type} &
  \multirow{2}{*}{Model} &
  \multirow{2}{*}{Training Data} &
  \multirow{2}{*}{Train/finetune Epochs} &
  \multirow{2}{*}{Model Size} &
  \multirow{2}{*}{Eval1} &
  \multicolumn{2}{c}{Eval2} \\ \cline{7-8} 
 &                     &                       &     &         &       & AEC + IVA & Spatialnet \\ \hline
\multirow{4}{*}{ASR Models} &
  Transformer &
  \multirow{4}{*}{\begin{tabular}[c]{@{}c@{}}190 hours \end{tabular}} &
  100 &
  29.89 M &
  31.75 &
  77.32 &
  58.23 \\
 & Conformer           &                       & 100 & 45.73 M & 26.89 & \textbf{69.55}   & 53.78      \\
 & E-Branchformer      &                       & 100 & 47.13 M & \textbf{26.05} & 71.04   & \textbf{51.52}      \\
 & Zipformer-Small     &                       & 100 & 30.22 M & 31.22 & 74.86   & 54.34      \\ \hline
\multirow{5}{*}{\begin{tabular}[c]{@{}c@{}}Open-Source\\ Models\end{tabular}} &
  Paraformer~\cite{gao2022paraformer} &
  60,000 hours &
  - &
  220 M &
  20.16 &
  74.04 &
  48.67 \\
 & Paraformer-Finetuned* & 190 hours & 10  & 220 M   & \textbf{16.65} & \textbf{66.68}   & \textbf{47.18}      \\
 & Whisper-Small~\cite{whisper}       & 680,000 hours         & -   & 244 M   & 50.69 & 79.49   & 65.72      \\
 & SenseVoice-Small~\cite{sensevoice}    & Over 400,000 hours    & -   & 234 M   & 24.63 & 75.58   & 50.64      \\
 & Qwen2-Audio~\cite{qwen2}         & 520,000 hours         & -   & 7B      & 29.92 & 76.24   & 54.48      \\ \hline
\end{tabular}
\end{table*}

\section{Baseline}

% https://github.com/MrSupW/ICMC-ASR\_Baseline
We develop a multi-channel in-car speech transcription system\footnote{https://github.com/DaiYvhang/AISHELL-5} based on the ICMC-ASR baseline\footnote{https://github.com/MrSupW/ICMC-ASR{\_}Baseline}. The baseline system consists of two primary sub-modules: speech frontend processing and automatic speech recognition (ASR). We provide the data preprocessing pipeline in the baseline system, and after processing the data, we train each sub-module independently.

During evaluation, Eval1 is directly processed by the ASR module. However, Eval2 undergoes a preprocessing step through the speech frontend module for echo cancellation and noise reduction before being fed into the ASR system. This process generates audio data for each speaker, which is subsequently segmented based on voice activity detection (VAD). Finally, the processed audio is passed through the ASR module to obtain the transcription. The training and inference process of our baseline is shown in Figure 1.

\subsection{In-car Speech Frontend Processing}
In our baseline system, we adopted a multi-channel acoustic echo cancellation (AEC) and an Independent Vector Analysis (IVA)~\cite{IVA} blind source separation algorithm. AEC is used to remove echo from the recorded microphone signal and IVA is used to separate speakers from different locations. For AEC, we use an adaptive filter to estimate and remove the echo from the microphone signal. Assuming that we have M microphones, the signal mixture can be represented as Eq.~\ref{eq1}. 
\begin{align}
\label{eq1}
    \mathbf{y}(t) = \mathbf{A} \mathbf{s}(t) + \mathbf{n}(t).
\end{align}
Here, $\mathit{t}$ is the time index, and $\mathbf{\textit{y}}(t)$ is the vector of the mixed signals received by the $\mathit{M}$ microphones. $\mathit{A}$ is an $\mathit{M} \times \mathit{N}$ mixing matrix, where $\mathbf{\textit{N}}$ is the number of speakers. $\mathbf{s}(t)$ represents the source signals of N speakers, and $\mathit{n}(t)$ is the noise vector.
In AISHELL-5, $M=4$ and $N=2$.
The goal of IVA is to separate the independent audio $\mathit{s}(t)$ for each seat from the mixed audio $\mathit{s}(t)$.
IVA is a maximum likelihood estimation (MLE) based blind source separation algorithm that separates the sources by minimizing the statistical dependence among the source signals. 
In IVA, assuming that the source signals are independent, the separation process can be achieved by maximizing the following objective function: 

\begin{align}
    \mathcal{L} = \sum_{t} \text{log}\left( \text{det}(\mathbf{A}(\mathbf{y}(t))) \right).
\end{align}

This objective function maximizes the statistical independence of the source signals to find the demixing matrix $\mathit{A}$, thus achieving the separation of the mixed signals.

In addition, our baseline system also integrates an end-to-end dereverberation, denoising, and separation approach based on Spatialnet, implemented using the NBSS\footnote{https://github.com/Audio-WestlakeU/NBSS}. Spatialnet makes extensive use of spatial information to perform multi-channel joint dereverberation, denoising, and separation, making it an ideal solution for in-car speech frontend processing. We use the same configuration as the open-source NBSS.

The model is trained using the complex mean squared error (cc\_mse) loss, with a loss scale factor of 100. To stabilize gradient computation, we use gradient clipping with a norm-based clipping algorithm and a predefined clipping threshold. The Adam optimizer is employed with a learning rate of 1e-3, and the learning rate scheduler follows an exponential decay strategy.

The training objective is to minimize the loss on the validation set. We monitor the model's performance using evaluation metrics such as Signal-to-Distortion Ratio (SDR), Scale-Invariant Signal-to-Distortion Ratio (SI-SDR), and Perceptual Evaluation of Speech Quality (PESQ).
The window length and hop size for STFT are 256 and 128, respectively. The real and imaginary parts of the 4-channel input audio are concatenated to form an 8-channel input feature. The near-end speech of the speaker at each seat is used as the training target. Additionally, 40 hours of real-recorded noise is mixed with the training data as an additional noise source.

Finally, we use a model trained for 100 epochs to perform frontend processing on Eval2.

\subsection{Automatic Speech Recognition}
To compare the performance of different ASR models on the test set, we evaluate several models, including Transformer~\cite{transformer}, Conformer~\cite{conformer}, E-Branchformer~\cite{ebranchformer}, and Zipformer~\cite{yao2023zipformer}, as baselines. The ASR models are trained using Wenet~\cite{wenet}, with Zipformer training and evaluation based on Icefall, though we only share the decoding results in the baseline system without providing the related code.

We use the AISHELL-5 training data for ASR training. For near-field data, we split the audio segments based on timestamp information. For far-field data, we first apply AEC and IVA for echo cancellation and speaker separation, then use VAD to segment the original long audio into shorter segments. This results in single-channel, single-speaker audio suitable for training. The final training data consists of near-field and high-quality far-field single-channel, single-speaker audio, totaling approximately 190 hours. During Zipformer training, the data is formatted to be compatible with the Icefall\footnote{https://github.com/k2-fsa/icefall}.

We use 80-dimensional fbank features as input, with utterance-level mean-variance normalization. The Adam optimizer is used, with a maximum of 100 training epochs. A warm-up and decay learning rate scheduler is employed, with a peak learning rate of 0.002, warm-up steps set to 25,000, and a batch size of 18. Finally, we average the last 30 epochs of all trained models and perform ASR inference on the averaged model, using attention rescoring for decoding with a beam size of 10.

\subsection{Evaluation and results}
We present the baseline system results on the test set, as shown in Table 3. The table displays results on the Eval1 task and cpCER results for the same ASR models on Eval2 after applying two different front-end processing methods. On Eval1, E-Branchformer achieved the best performance at 26.05\% with the same number of training epochs. On Eval2, ASR recognition results after Spatialnet processing showed a significant improvement, with about a 20\% gain, and E-Branchformer achieved 51.52\% cpCER. We also report results from some open-source models on the test set. Paraformer performed well among the open-source models, reaching 20.16\% on Eval1 and 48.67\% on Eval2 with Spatialnet as the front-end. After fine-tuning for 20 epochs with the same baseline training data, Paraformer improved these results to 16.65\% and 47.18\% with Spatialnet front-end processing.

\section{Conclusions}

This paper introduces AISHELL-5, currently the largest open-source multi-channel, multi-speaker free-talk in-car speech dataset, tailored for two key issues in current in-car speech processing techniques: the complex acoustic environment within the car and the frequent occurrence of overlapping speech from driver and passengers.
AISHELL-5 is suitable for speech separation, speech enhancement, noise reduction, and automatic speech recognition (ASR) tasks targeting in-car scenarios. 
All speech data is recorded from the real acoustic environment inside an electric vehicle, covering 60 different driving scenarios. 
It also includes 40-hour real-recorded noise data, which supports the in-car speech data simulation.
Moreover, we provide an open-access, reproducible baseline system based on this dataset.
% In addition, we have released a training and evaluation baseline. 
% In Eval1, the best result of our baseline is a CER of 26.05\%. In Eval2, we provided two different front- end processing strategies, AEC+IVA and Spatialnet, with the best performance achieving a cpCER of 37.75\%. 
We firmly believe that AISHELL-5, as the first open-source in-car multi-channel speech dataset, can offer a certain amount of support for this increasingly important ASR application scenario, advancing human-vehicle interaction towards greater accuracy and convenience.

\newpage

\ifinterspeechfinal
     
\else
     
\fi

\bibliographystyle{IEEEtran}
\bibliography{mybib}

\end{document}